\documentclass[12pt]{article}
\usepackage{amssymb}
\usepackage{graphicx}
\usepackage{amsmath}
\newcommand{\be}{\begin{equation}}
\newcommand{\ee}{\end{equation}}
\begin{document}
\baselineskip18pt
\title{L\'{e}vy flights and L\'{e}vy-Schr\"{o}dinger  semigroups}
\author{Piotr Garbaczewski\thanks{E-mail: pgar@uni.opole.pl}\\
Institute of Physics,  University  of Opole, 45-052 Opole, Poland}
\maketitle
\begin{abstract}

We   analyze two  different confining mechanisms for   L\'{e}vy
flights  in the presence of  external  potentials. One of them   is
due to  a  conservative  force  in the corresponding
Langevin equation. Another is implemented by
 L\'{e}vy-Schr\"{o}dinger semigroups   which induce
  so-called topological L\'{e}vy processes (L\'{e}vy flights with
  locally modified jump rates in the master equation).
Given a stationary  probability function (pdf) associated with the
Langevin-based fractional Fokker-Planck  equation, we demonstrate that  generically
 there exists a topological L\'{e}vy process with the very same
invariant pdf and in the  reverse.  
    \end{abstract}
     \noindent
 PACS numbers: 05.40.Jc, 02.50.Ey, 05.20.-y, 05.10.Gg
\vskip0.2cm

\section{Motivation}

L\'{e}vy flights stand for  a nickname  of a   class of symmetric  stable stochastic processes. These   non-Gaussian
  jump-type processes are not yet deeply accommodated  within  the general statistical physics  conceptual and technical
   imagery. 
One obstacle is rooted in purely technical reasons  -  a shortage of explicit analytic solutions,
e.g.  probability  density functions and transition probability densities.
Other obstacles may be attributed to  somewhat exotic  features of L\'{e}vy flights, like an existence of
  arbitrarily  small jumps or that of   fat tails  of the probability density  which  preclude  the existence of
   moments (in the least of the second one).

There is also a physically motivated obstacle:  no physical thermalization  mechanisms  have ever  been proposed  for
 L\'{e}vy  flights. Moreover, their physical "reason" (origin of noise)  appears to be  exterior  to the physical system,
  with no  reliable kinetic  theory background,   and therefore  no   fluctuation-dissipation response theory
  could have been   been set for  any stable noise.

  To the contrary, the noise  "reason"  is definitely an intrinsic feature of the
 environment-particle coupling in case of the standard  Brownian motion, based on the kinetic theory derivations.
 All traditional fluctuation-dissipation  relationships  find their place in the Brownian framework. None of them has been
  reproduced in the context of L\'{e}vy flights.

Nonetheless, L\'{e}vy flights remain an active research area, with a definite tendency to expand their field of applicability.
 A fraction of  papers, relevant to our problem  (taming L\'{e}vy flights via suitable confining potentials) is reproduced
 in the  reference list, \cite{levy}-\cite{dubkov}.

In the present paper we address an apparent   discord between two families of jump-type processes which can be  be associated
with the same type of the free (L\'{e}vy noise.
Namely, most of current research is devoted to Langevin equation-based derivations, where a deterministic force is perturbed
by  L\'{e}vy  (white) noise of interest, \cite{applebaum}-\cite{dubkov}.
However, in a number of publications, another class of  jump-type processes was  introduced under the name of topologically
induced super-diffusions, \cite{brockmann}-\cite{geisel1}. As we indicate in below  they belong to a broader class of
processes governed by L\'{e}vy-Schr\"{o}dinger  semigroups. In a  different  context,  they  were  introduced and investigated
 in some detail in Refs. \cite{klauder,olk},  c.f. also \cite{levy1}.

Fractional Fokker-Planck equations,  inferred respectively for  Langevin and topological
processes,  appear \it   not  \rm to  describe the same dynamical
pattern of behavior.  The statement of \cite{sokolov} was that
topologically induced fractional Fokker-Planck equation  does not
portray a situation equivalent  to any of  standard (e.g.
Langevin-based) fractional  Fokker-Planck equations and thus  does
not correspond to any  Langevin equation.

Confined L\'{e}vy flights have  received some attention in the recent literature \cite{fogedby}-\cite{dubkov}. On the
other hand, in the discussion of topological L\'{e}vy processes main emphasis has been put on their super-diffusive behavior
 with some neglect of  confining effects and the resultant  emergence of  non-Gibbsian  stationary probability
  densities, \cite{brockmann}-\cite{geisel1}.

We address the latter issue and set  general confinement criterions for    analytically
tractable of  Cauchy noise-driven processes. To this end, some ideas   have been
adopted  from the  general theory of diffusion-type stochastic   processes where
an  asymptotic approach towards  equilibrium (stationary density)  is one of
major topics of  interest, \cite{mackey}.

To handle topological L\'{e}vy processes we  invoke a theoretical
framework of the  Schr\"{o}dinger boundary data problem (whose
validity is  well established in the Brownian context,
\cite{zambrini,gar}). We shall  demonstrate that topologically
induced   processes of \cite{brockmann}-\cite{geisel1}  form a
subclass of its solutions with a properly tailored dynamical  semigroup and its (Feynman-Kac) potential, \cite{klauder,olk}.
That allows to take advantage of the   existing mathematical theory
of  L\'{e}vy processes and  L\'{e}vy-Schr\"{o}dinger semigroups, \cite{applebaum,sato} and
\cite{klauder,olk,olk1},  where  free  L\'{e}vy noise generators
  are additively perturbed by  suitable confining potentials. The theory  works well for both  Gaussian and non-Gaussian processes.
(For the record we indicate that,  in the Brownian case, the Schr\"{o}dinger problem incorporates  the   well known
 transformation of a Fokker-Planck equation into a generalized diffusion equation, \cite{risken}.)

In Section 2  we give a necessary  background for the forthcoming analysis of random motions in terms of dynamical
(Schr\"{o}dinger and next L\'{e}vy-Schr\"{o}dinger) semigroups.

 In Section 3 we  extend the Schr\"{o}dinger boundary data problem  Section 2 to L\'{e}vy flights. In Section 4 we   address
  in detail a number of model cases related to the Cauchy flights, with emphasis  on their confinement mechanisms.

\section{Brownian inspiration:  Schr\"{o}dinger semigroups}

\subsection{Gibbsian asymptotics of  Smoluchowski  processes}

We begin from a brief resume of
standard  Smoluchowski processes and their generalizations. Since
in the present discussion we assume  forward drifts to be of the
gradient form, our considerations  will be limited (albeit with no
substantial loss of generality) to one space dimension.

 Let  us consider $
\dot{x} = b(x,t) +A(t)$
with $\langle A(s)\rangle =0 \, , \,  \langle A(s)A(s')\rangle =
\sqrt{2D} \delta (s-s')$. Here, $b(x,t)$ is a forward drift of the
process, admitted to be  time-dependent,  unless we ultimately
pass to Smoluchowski diffusion processes where $b(x,t) \equiv
b(x)$ for all times and $x\in R$.

Given an initial   probability density $\rho_0(x)$. We know that
the diffusion process drives this density in accordance with
 the  Fokker-Planck equation
$
\partial _t\rho = D\triangle \rho -  \nabla \, ( b \cdot \rho ) \, .
$
We introduce an osmotic  velocity  field  $u = D\nabla \ln \rho $,
together with the current velocity field  $v=b - u$. The latter
obeys  the continuity equation $\partial _t \rho = - \nabla j$,
where  $j= v\cdot \rho $  has   a standard interpretation of a
probability current.

Presently we   pass to time-independent drifts of the  diffusion process,
 that are induced by   external (conservative,  Newtonian) force fields $f= - \nabla V$.
One arrives at Smoluchowski diffusion processes by  setting
 \begin{equation}
b = {\frac{f}{m\beta }} = - {\frac{1}{m\beta }}  \nabla V \, .
\end{equation}
This  expression accounts for the fully-fledged phase-space derivation of the spatial process, in the regime of large $\beta $.
 It is  taken  for granted that  the fluctuation-dissipation
  balance gives rise to the  standard form $D=k_BT/m\beta $ of the  diffusion coefficient.

Let us consider a stationary asymptotic  regime, where $j\rightarrow j_*=0$.
We denote an  (a priori assumed to exist, \cite{mackey}), invariant density    $\rho _*= \rho _*(x) $.
Since   $v_*=0$, there holds
\begin{equation}
b_*=u_* = D \nabla  \ln \rho _*  \, .
\end{equation}
 However    $b=f/m\beta  $  by its very definition  does  not  not depend functionally
 on the probability density. Thus
$b= b_*$  and therefore $
  \rho _*(x) = (1/Z)\,  \exp[ - V(x)/k_BT]$.
Our outcome   has   the    Gibbs-Boltzmann  form.

Denoting  $  F_{*}\doteq  - k_BT
\ln Z $, where  $Z= \int \exp(-V/k_BT) \, dx$, we have
\begin{equation}
\rho _*(x) =
\exp\left( [F_* - V(x)]/k_BT\right) \doteq \exp [2\Phi (x)]\, .
\end{equation}
Here, to comply with the notation of
\cite{zambrini,klauder} and the forthcoming discussion of
a  topological generalization of the Brownian motion and next
L\'{e}vy flights  \cite{brockmann}-\cite{geisel1}, we have
introduced a new potential function $\Phi $ such that $\rho _*^{1/2}
= \exp \Phi $ and  $b= 2D \nabla \Phi $.

\subsection{Schr\"{o}dinger semigroups and their integral  kernels}

Given Eq. (3), let us  re-define a probability density of  the Smoluchowski process as follows
\begin{equation}
\rho (x,t) \doteq  \theta _* (x,t) \exp [\Phi (x)]\, .
\end{equation}
This  allows   to map a Fokker-Planck equation into an
associated generalized heat equation \cite{risken}.
Indeed,  if  $\rho (x,t)$ obeys the Fokker-Planck equation whose
 drift $b(x)$ has the gradient form, the inferred $\theta _*(x,t)$
obeys the parabolic   equation
\begin{equation}
\partial _t\theta _*= D \Delta \theta _* - {\cal{V}} \theta _* \, ,
\end{equation}
provided a compatibility  condition
 \begin{equation}
 {\cal{V}}(x) = {\frac{1}{2}}\left( {\frac{b^2}{2D}} + \nabla b\right)
\end{equation}
holds true.

We have here arrived at  a contractive  semigroup dynamics
$  \theta _*(t) =  [ \exp(-t\hat{H})\theta _*](0)$,
with  the  dynamical  (Schr\"{o}dinger) semigroup operator
$\exp(-t\hat{H})$ whose generator  $-\hat{H}$ involves the familiar
(here, $1/2mD$ rescaled) Schr\"{o}dinger Hamiltonian
\begin{equation}
\hat{H} = -D\Delta + {\cal{V}} \, .
\end{equation}
The initial $t=0$ data read  $\theta _*(x,0)$.

For ${\cal{V}}={\cal{V}}(x), x\in R$,  typically expected to be continuous  and  bounded from below, the
operator  $\hat{H}=-D \triangle + {\cal{V}}$ is essentially
self-adjoint on a natural dense subset of $L^2$, and the  integral
kernel ($s<t$)
\begin{equation}
k(y,s,x,t)=\left( \exp[-(t-s)\hat{H}] \right) (y,x)
\end{equation}
of the dynamical semigroup $\exp(-t\hat{H})$  is strictly positive. Assumptions concerning an
  admissible potential may be relaxed. The necessary demands are that $\hat{H}$ is self-adjoint
  and bounded from below.
Once we have an integral kernel, the semigroup dynamics takes the form
$\theta _* (x,t) = \int \theta _*(y,s)\, k(y,s,x,t)$.

 The key role,  in the transformation of the
Fokker-Planck operator  and equation into  Schr\"{o}dinger-type operator and (parabolic) equation,
   is  played by the  potential  ${\cal{V}} $.
We recall that
 $-D\Delta $ is a  semigroup  generator for  the free Brownian
motion which yields the heat equation $\partial _t \rho = D\Delta \rho $.

Brownian  noise perturbations  of   conservative force fields  may  be encoded on two equivalent ways.
One approach directly  employs the Langevin equation, c.f our discussion of Smoluchowski processes.
It is the deterministic force field $\sim -\nabla V$ which one additively
perturbs by  white noise.

Another  approach is based on
the classic Feynman-Kac formula which uniquely   assigns a    dynamical  semigroup kernel  (and dynamics)  to
a  concrete   potential ${\cal{V}}(x)$:
\begin{equation}
k(y,s,x,t)= \int exp\bigl [- \int_s^t {\cal{V}} \bigl (X(u),u\bigr
)du \bigr] \, d\mu [s,y\mid t,x] \, .
\end{equation}
In the above we integrate the  $exp[- \int_s^t {\cal{V}} (X(u),u)
du]$ weighting factor with respect to the conditional Wiener measure
i.e. along all sample paths (Brownian trajectories)  of the Wiener
process  $X(t)$ which connect $y$ with $x$ in time $t-s$;  $X(t)$
stands for the random variable of the Wiener process.

\subsection{Schr\"{o}dinger's boundary data problem}

Schr\"{o}dinger semigroups  naturally appear in the framework of so-called Schr\"{o}dinger boundary
data problem, \cite{zambrini}.
We assume to   have   a priori  given  two probability densities
$\rho (x,0)$ and $\rho (x,T)$, $T>0$ and  attempt to construct a
Markov process  that interpolates between these boundary data.

The major  outside  input  is  a definite choice of a bounded  strictly positive and continuous
  integral   kernel  $k(x,s,y,t)$,  for all times in the interval (possibly infinite) $ 0\leq s<t\leq T$.
  Our   considerations  are restricted  to
  Feynman-Kac  kernels and so to  Schr\"{o}dinger  semigroups. Each semigroup is selected  by prescribing
  the  Feynman-Kac  potential  ${\cal{V}}$, c.f.  Eqs. (7) and  (9),  which needs to be a continuous function.
  For each  concrete  choice of the semigroup  kernel, cf. \cite{zambrini,acta}.

By denoting  (notice that we have here combined both forward and
backward propagation scenarios in the time interval $[0,T]$):
\begin{equation}
\theta _*(x,t)=\int f(z) k(z,0,x,t) dz, 
\end{equation}
$$\theta (x,t)= \int k(x,t,z,T)g(z) dz$$
where functions $f(x)$ and $g(x)$ are suitable inital/terminal data, 
 it follows that for all $0\leq s<t\leq T$ there holds
\begin{equation}
\rho (x,t) \doteq  \theta (x,t)\theta _*(x,t) = \int p(y,s,x,t) \rho
(y,s) dy \, .
\end{equation}
Here
\begin{equation}
p(y,s,x,t) = k(y,s,x,t){{\theta (x,t)}\over \theta (y,s)}
\end{equation}
is the transition probability density of the pertinent Markov
process that interpolates between $\rho(x,0)$ and $\rho (x,T)$.
 We note that  $\theta _*(x,0)= f(x)$ while $\theta (x,T)=g(x)$.

 In case of the free evolution,  by  setting $\theta (x,t)= \theta \equiv const $,   we effectively  transform
   an integral kernel $k$  of the $L^1(R)$   norm-preserving semigroup  into a transition probability density $p$ of the
    Markov stochastic process, e.g. the familiar heat kernel.

      Our $\theta $ and $\theta _* $
    are defined up to constant factors. This  freedom is limited by the demand that  $\theta \cdot \theta _*\doteq \rho $  actually
     is   a probability density of the pertinent Markov process.

     We  emphasize that in addition to the semigroup dynamics, to secure the uniqueness of an
      interpolating Markov  process,    we need an auxiliary motion rule, c.f. \cite{zambrini}, (propagating  backwards
      in time the  prescribed terminal function $\theta (x,T)$)
     \begin{equation}
\partial _t\theta  = - D \Delta \theta + {\cal{V}} \theta \, .
     \end{equation}
For  time-independent $\theta = \theta (x) = \exp \Phi (x)$ we note that ${\cal{V}} = D\, \Delta \theta /\theta $.
 Taking into account the definition  of the forward drift
       $b= 2D \nabla \theta / \theta $, \cite{zambrini},  we end up with  the previous compatibility condition  (6).

 In case of Smoluchowski processes we encounter  asymptotic invariant densities,
\cite{mackey}. Accordingly, $\theta \sim \rho ^{1/2}_*$ and thus
\begin{equation} {\cal{V}} = D \,
{\frac{ \Delta \rho _*^{1/2}}{\rho _*^{1/2}}}  \, .
\end{equation}
This result derives from the semigroup argument alone and does not rely on the Gibbsian form of $\rho _*$.

\section{L\'{e}vy-Schr\"{o}dinger  semigroups}

\subsection{Stable noise and its generator}

The  Schr\"{o}dinger  boundary data problem is amenable to an immediate  generalization to infinitely divisible probability laws which
 induce   contractive semigroups  (and their  kernels)  for   general  Gaussian and  non-Gaussian noise models,
 \cite{klauder} and \cite{olk}-\cite{cufaro}.  
 A subclass  of stable probability laws  contains a
  subset that is   associated in the literature  with the concept of   L\'{e}vy flights.
  
Instead of the  semigroup generators  proper,  we  shall  employ  the   rescaled   Hamiltonians (true generators have an opposite sign)
of the form $\hat{H}=F(\hat{p})$, where  $\hat{p}=-i \nabla $ stands for the momentum  operator  (up to the scaled away  $\hbar $ or $2mD$ factor),
 and  for $-\infty <k<+\infty $,  the function  $F=F(k)$ is   real valued,
bounded from below and  locally integrable.
 Then, for a function $f(x)$ in the domain of the semigroup operator, we have 
\begin{equation}
[exp(-t\hat{H})f](x)= [exp(-tF(p)) \tilde{f}(p)]^{\vee }(x)
\end{equation}
where the superscript $\vee $ denotes the inverse Fourier transform, and  $\tilde{f}$ stands for   the Fourier transform of $f$.
Let us set
\begin{equation}
k_t={1\over {\sqrt {2\pi }}}[exp(-tF(p)]^{\vee }\, .
\end{equation}
The
action of $exp(-t\hat{H})$ can be given in terms of a convolution (i.e. by means of an integral kernel $k_t\equiv  k(x-y,t)=k(y,0,x,t) $):
\begin{equation}
exp(-t\hat{H})f = [\exp(-tF(p))\tilde{f}(p)]^{\vee } = f*k_t
\end{equation}
where
\begin{equation}
(f*g)(x): =\int_R g(x-z)f(z)dz \, .
\end{equation}

 We shall restrict considerations only to those $F(p)$ which give rise to
 positivity preserving semigroups:  if $F(p)$ satisfies the celebrated
 L\'{e}vy-Khintchine formula, then $k_t$ is a positive measure for all
 $t\geq 0$.
The most general case refers to a  combined  contribution from three types of
processes:  deterministic, Gaussian, and  the jump-type process.

We recall that a characteristic function of a random variable $X$  completely determines a probability distribution of that variable.
If this distribution admits a density we can write $E[\exp(ipX)] =  \int_R \rho (x) \exp(ipx) dx$ which,
  for infinitely divisible probability laws,  gives rise to:
\begin{equation}
{F(p) = -   \int_{-\infty }^{+\infty } [exp(ipy) - 1 -
{ipy\over {1+y^2}}]
\nu (dy)}
\end{equation}
where $\nu (dy)$ stands for the appropriate  L\'{e}vy measure. The corresponding non-Gaussian Markov process is characterized by
\begin{equation}
E[\exp(ipX_t)]= \exp[-t F(p)] \, .
\end{equation}
  Accordingly, the contractive  semigroup generator  follows  (keep the minus sign in memory) from: $F(\hat{p})= \hat{H}$.

For  the sake of clarity   we restrict further  considerations to   non-Gaussian random variables whose probability densities
are centered and symmetric, e.g.  a subclass of stable distributions characterized by
\begin{equation}
F(p) = \lambda   |p|^{\mu } \Rightarrow \hat{H} \doteq   \lambda |\Delta |^{\mu /2} \, .
\end{equation}
Here  $\mu <2$ and $\lambda >0$ stands for the intensity parameter
 of the L\'{e}vy  process. To account for the interval  $0\leq \mu \leq 2$ boundaries,  one should rather
  employ $(-\Delta )^{\mu /2}$ instead of $|\Delta |^{\mu /2}$, since   $-\Delta $ is a positive operator.

The  fractional Hamiltonian   $\hat{H}$, which is a pseudo-differential operator,  by construction  is  positive
 and self-adjoint on a properly tailored  domain.  A  sufficient and necessary
   condition for both these  properties
  to hold true is that the probability density of the  L\'{e}vy process is symmetric, \cite{applebaum}.

The associated  jump-type   dynamics is interpreted  in terms of  L\'{e}vy flights. In particular
\begin{equation}
F(p)= \lambda  |p| \rightarrow \hat{H}= F(\hat{p}) =  \lambda |\nabla | \doteq \lambda (-\Delta )^{1/2}
\end{equation}
refers to the Cauchy process, see e.g. \cite{klauder,olk,olk1}.

The pseudo-differential  Fokker-Planck equation, which  corresponds to the fractional Hamiltonian  (22) and the
fractional semigroup $\exp(-t\hat{H}_{\mu })=\exp(-\lambda |\Delta |^{\mu /2})$, reads
\begin{equation}
\partial _t \rho  = -  \lambda |\Delta |^{\mu /2} \rho  \, ,
\end{equation}
to be compared with the heat equation $\partial _t \rho  = +  D \Delta  \rho $.

\subsection{Response to external potentials: stationary densities}

 \subsubsection{Langevin scenario}

In case of jump-type (L\'{e}vy) processes a response to external perturbations by conservative force fields appears to be
 particularly interesting.
 On the one hand, one encounters  a widely accepted reasoning, \cite{fogedby}-\cite{dubkov} where   the   Langevin equation,
with additive   deterministic   and  L\'{e}vy "white noise"  terms,  is found to imply a  fractional Fokker-Planck equation,
 whose form  faithfully  mimics  the Brownian  version, e.g. (c.f. \cite{fogedby}, see also \cite{olk1})
 \begin{equation}
\dot{x}= b(x)  + A^{\mu }(t) \Longrightarrow \partial _t\rho = -\nabla (b\cdot \rho ) - \lambda |\Delta |^{\mu /2}\rho
 \end{equation}
where we keep the notation  $b= f/m\beta  $, $f=-\nabla V$ of Eq. (1).

 We emphasize a difference in sign in the
second term, if compared with Eq. (4) of Ref. \cite{fogedby}.  There, the minus sign is  absorbed in the adopted definition
of the (Riesz) fractional derivative. Apart from the formal resemblance of operator symbols, we do not directly
 employ fractional derivatives in our formalism.

The validity of (24) and temporal details of  an approach towards an asymptotic invariant density were investigated  for the
Cauchy-Ornstein-Uhlenbeck proces \cite{olk1}.  That  safely extends  to more general  stable (symmetric)
OU  processes, \cite{fogedby} whose asymptotic behavior directly  follows from the Fourier transform of $\rho (x,t)$.

In contrast to the standard Gaussian case (folk theorem: all solutions of a given Fokker-Planck equation have a
 common asymptotic behavior), there are no  general   mathematically rigorous  statements about the asymptotic behavior
  of solutions  of a general  fractional
Fokker-Planck equation.  Specifically, this comment pertains to whether (if at all) and  how  (detailed convergence estimates)
 invariant (stationary, equilibrium)  densities are  asymptotically approached   in the course of  a  nonlinear (symmetric)
  stable process.
  Anyway, for  monomial  drifts, analytic forms of associated invariant densities  were explicitly elaborated
in the presence of  Cauchy noise in Refs. \cite{chechkin,dubkov0,dubkov}.

 It is also  well known that, given the drift $b$ potential  $V(x)$, (1), an  invariant density for a confined L\'{e}vy flight
  does not  show any connection with the Gibbsian exponential  of the form (3).   Hence,
   the usefulness of the transformation (4) comes under scrutiny in the
presence of L\'{e}vy noise.

\subsubsection{Feynman-Kac (topological)  route}

It is a possible  to account for   external perturbations by means  of L\'{e}vy-Schr\"{o}dinger Hamiltonians,
\cite{cufaro} and \cite{klauder,olk} where  a potential function ${\cal{V}}(x)$ appears as a necessary ingredient.
Assuming that its  functional form  guarantees that
\begin{equation}
\hat{H}_{\mu }  \doteq   \lambda |\Delta |^{\mu /2} +  {\cal{V}}\, .
\end{equation}
is self-adjoint and positive in a suitable Hilbert space, we can consistently   introduce the
 L\'{e}vy-Schr\"{o}dinger semigroup $\exp(- t\hat{H}_{\mu })$
 and  the  fractional  relative of the
generalized diffusion equation:
\begin{equation}
\partial _t\theta _* = -  \lambda |\Delta |^{\mu /2} \theta _*   -  {\cal{V}} \theta _*   \, .
\end{equation}
The time-adjoint equation  (compare e.g. (15)) has the form
\begin{equation}
\partial _t\theta = \lambda |\Delta |^{\mu /2} \theta    + {\cal{V}} \theta    \, .
\end{equation}
 Clearly, we have here reproduced the general  theoretical framework of the  Schr\"{o}dinger boundary data problem, where
    $\theta ^*(x,t) \theta (x,t) =\rho (x,t)$ stands for a   probability density of an associated  Markov process.

Let   $\rho _*(x)$  be a stationary  (invariant, equilibrium) probability
density of the pertinent  process.
With  $ \theta (x,t) \equiv \theta (x)= \exp[\Phi (x)]$, we can  mimic the trial decomposition ansatz,  Eq. (4)
 $\theta _* = \rho \, \exp (- \Phi )$.

If we set  $\exp[ \Phi (x)] = \rho ^{1/2}_*(x)$, then   Eq. (27)   takes the form of the compatibility condition,
akin  to that of Eq. (14):
\begin{equation}
{\cal{V}}  =   -\lambda\,  {\frac{|\Delta |^{\mu /2}\,  \rho ^{1/2}_*}{\rho ^{1/2}_*}} \, .
\end{equation}
 This identity should be  compared with Eq. (8) in Ref. \cite{geisel},  where an analogous   effective potential (up to a systematic
  sign difference)  was deduced for  the fractional  L\'{e}vy-Schr\"{o}dinger  type  equation,  in the study of L\'{e}vy flights in inhomogeneous media.

  In view of (26) and (27) we have a continuity equation with an explicit fractional input
 \begin{equation}
  \partial _t \rho  = \theta \partial _t \theta ^*= -   \lambda  ( \exp \Phi ) |\Delta |^{\mu /2}[ \exp(-\Phi )
    \rho ]
    - {\cal{V}} \cdot \rho  \doteq - \nabla j \, .
 \end{equation}
Let us  make  cosmetic changes $\Phi \rightarrow -V/2k_BT$, c.f.  Eqs. (3) and (17) for comparison). Next we
take advantage of   $\rho = \theta _*\theta $ with $\theta = \exp (-\beta V/2) $  and  Eq. (28).
By setting  $\lambda =1$ and  $\beta = 1/k_BT$, we  give   Eq. (35)   a familiar  form of the  transport equation  previously
  introduced  in  a number of papers:
\begin{equation}
\partial _t \rho =-    \exp(-\beta V/2)\,  |\Delta |^{\mu /2} \exp(\beta V/2 )  \rho  +
\rho \exp (\beta V/2) |\Delta |^{\mu /2} \exp(-\beta V/2) \, ,
\end{equation}
 c.f. formula (6) in \cite{geisel}, formula  (5) in \cite{geisel1} and  formula  (36) in \cite{brockmann}.
There,   the investigated process was named a topologically induced super-diffusion.  We point out  a systematic  sign  difference
between  our  $|\Delta |^{\mu /2}$   and the corresponding fractional derivative  $\Delta ^{\mu /2}$
   of \cite{brockmann,geisel,geisel1}. Graphically  these symbols look similar, but have different origin.

The  major observation at this point is, that  topological L\'{e}vy (specifically, Cauchy) processes have been embedded
 into a well developed mathematical framework of Ref.
\cite{olk}. Therefore,   we can reconcile  any specific choice of $\theta \sim \rho _*^{1/2}$ with  minimal requirements  upon the
properties of ${\cal{V}}$, Eq. (28), that guarantees  the existence of  L\'{e}vy-Schr\"{o}dinger  semigroup  and thence of
the fractional transport equations (29), (30).

\subsubsection{A discord and its analysis}

The puzzling point, raised in \cite{brockmann},  is that for  L\'{e}vy processes in external force fields,
the Langevin approach yields a continuity
(e.g. fractional Fokker-Planck) equation in a  form
\begin{equation}
\partial _t\rho = -\nabla \left(- {\frac{\nabla V }{m\beta }}\, \rho  \right) - \lambda |\Delta |^{\mu /2}\rho
 \end{equation}
 that is very different from the previous fractional transport equation (29).

The conclusion of Refs. \cite{brockmann}-\cite{geisel1} was   that,  while  assuming   $\Phi \sim  -V$
where $V$ is the external force potential (up
 to inessential factors),  the two transport equations (29) and (31) are  plainly  incompatible. Eq.  (31) seems not
  to correspond to any standard  Langevin equation  with  L\'{e}vy noise term and $b=-\nabla V/m\beta$ as  its
   deterministic part  and in reverse.

 Apart from the verbal statement this puzzling discrepancy has not been explored in more depth. We shall partly fill this gap in below.

  Let us take  the fractional Fokker-Planck equation (31)  for granted.
   Assume that   $\rho (x,t)\rightarrow  \rho _*(x)$, as  $t\rightarrow \infty $.
  Let   the  invariant probability
    density of the fractional equation  (31) determines  $\Phi $ through $\exp[ \Phi (x)] = \rho ^{1/2}_*(x)$.
    We know that in case of L\'{e}vy
    noise  $\rho _* (x)$ \it  does not \rm have a Gibbsian structure $\rho _* \sim \exp (- V/k_BT)$, with $V$ being the drift potential.

The general problem to be addressed is:\\
 (i) choose a functional form of $V(x)$ and thus the drift of the Langevin-type  process,\\
  (ii) infer  an invariant density
 $\rho _*$ that is compatible  with the  fractional  Fokker-Planck equation  (31),\\
  (iii) given $\rho _*$,  deduce  the  corresponding  Feynman-Kac (e.g. dynamical semigroup)  potential ${\cal{V}}$
  by means of (28): the two dynamical scenarios (29) and (31) would thus share a common stationary density,\\
(iv) use  ${\cal{V}}$ in (28),(29)  and  verify whether  the  "topologically induced
 dynamics" is  affine, if at all,   to that associated with (31) (and thus to  the underlying
 Langevin equation  with L\'{e}vy noise),\\
 (v) check an asymptotic behavior  of $\rho(x,t)$ in both  scenarios (29) and (31) to find possible differences
 in the speed (convergence  time rate) with which the common invariant density  $\rho _*(x)$ is   approached.\\
(vi) repeat the procedure in reverse  by starting from (iii) and then deduce the
drift for the Langevin equation with L\'{e}vy noise; next  compare the dynamical scenarios (29) and (31)
 for any common initial probability density.

In below, we shall  mostly  concentrate on the above points (i)-(iii). However,  their  reverse (vi) will  receive some
 attention as  well.  We note  that problems (iv)-(vi) need more detailed computer-assisted analysis, that is postponed to a future publication.

\section{Processes induced by Cauchy noise: Invariant density  vs   semigroup (Feynman-Kac)  potential}

In view of  serious  technical difficulties,   we shall  not attempt a fully fledged solution to the just formulated problem,
for any symmetric  stable  jump-type process and any conceivable drift.
Instead, we turn our attention to situations where explicit functional forms of invariant densities are
available.

 Most of them were inferred in connection with Cauchy noise, \cite{klauder,olk1}, \cite{fogedby}-\cite{dubkov}.
In particular, attention has been paid to confining properties of various  monomial  drifts  upon the Cauchy noise. On the other hand,
 L\'{e}vy flights through a  "potential landscape" (topological processes of \cite{brockmann}-\cite{geisel1}) were
 interpreted as (enhanced) super-diffusions and an issue of possible asymptotic densities has not been significantly developed.

For a pseudo-differential operator $|\Delta |^{\mu /2}$,
    the action on a function from  its domain is  greatly simplified, in view of the properties of the  L\'{e}vy
    measure $\nu _{\mu }(dx)$.
We have  \cite{klauder,sokolov,olk,cufaro,dubkov}:
$$
(|\Delta |^{\mu /2} f)(x)\, =\, - \int_R [f(x+y) - f(x) - {{y\, \nabla f(x)}
\over {1+y^2}}]\, \nu _{\mu }(dy)
$$
$$\Downarrow $$
\begin{equation}
 (|\Delta |^{\mu /2} f)(x)\, =\, - \int  [f(x+y) - f(x) ] \nu _{\mu }(dy) \, .
\end{equation}
The Cauchy-L\'{e}vy measure,   associated with the  Cauchy  semigroup generator $|\Delta |^{1/2}\doteq |\nabla |$, reads
\begin{equation}
\nu  _{1/2}(dy) = {\frac{1}{\pi }} {\frac{dy}{y^2}}\,
\end{equation}

By changing  an integration variable $y\rightarrow z=x+y$, we give  Eq. (38)  the familiar form
\begin{equation}
 (|\nabla | f)(x)\, =\, -  {\frac{1}{\pi }} \int  {\frac{f(z)- f(x)}{|z-x|^2}}\,  dz
\end{equation}
where $1/\pi |z-x|^2$ has an interpretation of an intensity with which jumps of the size $|z-x|$ occur. Once we set $f = \exp \Phi $,
 the "topologically induced" jump-type process of Refs. \cite{brockmann}-\cite{geisel1} is in fact obtained.

  \subsection{Ornstein-Uhlenbeck-Cauchy process}

In case of the Ornstein-Uhlenbeck-Cauchy (OUC) process, the drift is given by $b(x)= - \gamma x$, and an asymptotic
invariant density associated with
\begin{equation}
\partial _t \rho = - \lambda |\nabla | \rho + \nabla [(\gamma x)\rho ]
\end{equation}
 reads:
\begin{equation}
\rho _*(x) = {\frac{\sigma }{\pi (\sigma ^2 + x^2)}}
\end{equation}
where $\sigma = \lambda /\gamma $, c.f. Eq. (9) in Ref. \cite{olk1}.

Note that a characteristic function of this density
reads $-F(p) = - \sigma  |p|$ and gives account of a    non-thermal fluctuation-dissipation balance.
The modified noise intensity parameter $\sigma $ is  a ratio of
 an intensity parameter  $\lambda $  of the  free
  Cauchy noise and  of the friction coefficient  $\gamma $.

   For the Cauchy random variable   $X_t$  we have
 $E[\exp (ip X_t)] = \exp( t\lambda |p|)$ where $\lambda $ stands for an intensity of the  Cauchy process.
The corresponding  (time-dependent) probability density has the form (36) with $\sigma  \sim  t\lambda $, e. g. $\rho (x,t)= \lambda t/\pi
 [(\lambda t)^2 + x^2]$.

Here,   $\sigma $ and likewise  $t\lambda $  play the role of  scale parameters  which specify
the half-width  of the  Cauchy density at its half-maximum.   Since $t\lambda $ grows monotonically, the corresponding free Cauchy
 noise  probability density  flattens and its maximum  drops down with the flow of  time.

In view of  $\sigma = \lambda / \gamma $,  the frictional drift  $-\gamma x$  may stop the "flattening"   of the probability distribution
  and stabilize  the density  at quite arbitrary  shape  (with respect to its maximum and  half-maximum related half-width),   by
 manipulating   $\gamma $. For example, $\gamma \gg 1$  would   induce a significant shrinking  of
 the distribution $\rho _*$, if compared to the  reference (free noise)  probability density at any  time $t \sim 1/\lambda $.
 In parallel, a  maximum value of the density   would increase:
$1/\pi \lambda \rightarrow \gamma / \pi \lambda $.

Clearly, large  friction has a confining effect on Cauchy noise. Confined L\'{e}vy flights, and specifically confined
  Cauchy flights, were analyzed  before  in \cite{chechkin}-\cite{dubkov} with the aim to produce explicit
    invariant probability  densities in
the presence of external  (confining)  forces. Their properties proved to be  quite interesting
  for monomial and polynomial choices of $V(x)$.

To deduce the Feynman-Kac potential ${\cal{V}}$ for the OUC   process, given an invariant
 density $\rho _*$, Eq. (36), we need to evaluate
 \begin{equation}
{\cal{V}}(x)  = {\frac{\lambda }{\pi }} (\sigma ^2 + x^2)^{1/2}  \int \left[ {\frac{1}{\sqrt{\sigma ^2 + (x+y)^2}}} -
{\frac{1}{\sqrt{\sigma ^2 + x^2}}} \right] {\frac{dy}{y^2}} \, .
 \end{equation}

In the  notation $a=\sigma ^2 + x^2$,  $b= 2x$, $R(y)= \sigma ^2 +  (x+ y)^2$  the indefinite integral reads, \cite{gradstein}:
\begin{equation}
{\frac{\lambda }{\pi }}\,   \int\left[
 {\frac{\sqrt{a}}{y^2\sqrt{R(y)}}}  -  {\frac{1}{y^2}} \right] dy = {\frac{\lambda }{\pi }} \left[ -   {\frac{\sqrt{R(y)}}{y \sqrt{a}}}
 + {\frac{b}{2a}}\,  Arsh \left(
 {\frac{2a+by}{2\sigma |y|}}\right)
 + {\frac{1}{y}}\right]\, .
\end{equation}
Because of the singularity at $y=0$, we  must handle  the integral in terms of its principal value, i.e. by resorting to
$\int \rightarrow \int_{-\infty }^{-\epsilon } + \int_{\epsilon }^{+\infty }$, and  next  performing the $\epsilon \rightarrow 0$ limit.

Taking into account that $Arsh \, x \equiv  \ln (x+ \sqrt{1+x^2})$, \cite{stephanovich},  we ultimately  get
\begin{equation}
{\cal{V}}(x) = {\frac{\lambda }{\pi }} \left[ - {\frac{2}{\sqrt{a}}} + {\frac{x}{a}}\ln {\frac{\sqrt{a}+x}{\sqrt{a}-x}}\right] \, .
\end{equation}
${\cal{V}}(x)$  is bounded both from below and above, with the asymptotics  $(2/|x|) \ln |x|$ at infinities, well fitting to the
 general mathematical construction  of (topological) Cauchy processes in external potentials, \cite{olk}.

  Accordingly, we  know for sure that there exists  a topological
 Cauchy process  with the Feyman-Kac potential ${\cal{V}}$, Eq. (39), whose invariant density coincides with that
 for the Langevin-supported  OUC process.

 At the moment, we have nothing to say about a detailed time-dependent behavior of the topological process and
  a particular scenario of an approach towards  the invariant density  (equilibrium) in the large time regime.
The two considered  jump-type processes, whose dynamics is embodied respectively   in the fractional Fokker-Planck equation and
in the  L\'{e}vy-Schr\"{o}dinger  semigroup (topological case) dynamics  definitely  stay in affinity, since they share a
 common invariant density.
In the  near-equilibrium regime, any dynamical distinction between these motion scenarios, becomes immaterial.
However, their  detailed  dynamical  behavior far-from-equilibrium  might be different and this issue deserves further exploration.

There is no  faithful  Langevin-type representation of a topological process and in reverse, even though an invariant density is
common for both.  Nonetheless, we have demonstrated that by staring from a common initial probability density, the
two  (Langevin and dynamical semigroup) motion scenarios  may, in principle,   end up at a common invariant density.

\subsection{Confined  "topological" Cauchy  process}

Neither the  OUC process  nor its topological relative are confined. For the Cauchy  density, the  second moment is nonexistent.
We shall verify  the outcome of the OUC discussion for Cauchy-type processes whose invariant densities admit  the second  moment.
Let us   consider the quadratic Cauchy density:
\begin{equation}
\rho _*(x) = {\frac{2}{\pi }}\, {\frac{1}{(1+x^2)^2}}\, .
\end{equation}
The action of $|\nabla |$ upon this density can be  evaluated
by recourse to the free Cauchy  evolution.

We note  that  $  (1/\sqrt{2\pi })\rho_*^{1/2} =
(1/\pi )/(1+x^2)$  actually is the Cauchy probability  density.
Let us  consider $f(x) = \rho _*^{1/2} $ as the initial data for
the free Cauchy evolution $\partial _t f = \lambda  |\nabla |f$. This   takes $f(x)$ into
\begin{equation}
f(x,t)= {\frac{2}{\pi }} \, {\frac{1 + \lambda t}{[(1 + \lambda t)^2 + x^2]}} \, .
\end{equation}
Since
\begin{equation}
 \lambda |\nabla |f =  - \lim_{t\downarrow 0} \partial _t f
\end{equation}
we end up with
\begin{equation}
{\cal{V}} (x) = {\frac{\lim_{t\downarrow 0} \partial _t f}{f}}(x) = \lambda  {\frac{x^2-1}{x^2 +1 }}\, .
\end{equation}
The shape of this potential is quite inspiring.  A minimum $ -\lambda $ is achieved at $x=0$, ${\cal{V}}=0$  occurs for $x=\pm 1$,
a maximum $+\lambda $ is reached at $x \rightarrow  \pm \infty $.

The potential is bounded both from below and above, hence can trivially be made non-negative (add $\lambda $).
 Therefore,  the invariant  density  (40) is  fully  compatible with the general construction of
 the Cauchy-Schr\"{o}dinger semigroup and the induced jump-type process,  c.f. Corollary 2, pp. 1071 in \cite{olk}.
  This topological Cauchy  process is
  induced by the Cauchy generator plus a potential function ${\cal{V}}$ given by Eq. (43).
   c.f. Corollary 2, pp. 1071 in \cite{olk}.
 The process is of the  jump-type  and  can be obtained as an $\epsilon \downarrow 0$ limit of a step process, e.g. jump process
 whose jump size is bounded from below  by $\epsilon >0$ but unbounded from above.

In connection with the problem (vi) o Section 3.2.3 let us note that if  the  quadratic Cauchy  density  (40) would  actually stand for
  a stationary density   of  the  fractional Fokker-Planck equation  with a drift Eq. (31), then  we should  have:
  \begin{equation}
\partial _t \rho _* = 0 = - \nabla (b\, \rho _*) - \gamma |\nabla |\rho _* \, .
  \end{equation}
Therefore the drift function, if any,  may be  deduced by means of an indefinite integral:
\begin{equation}
b(x) = -  {\frac{\gamma }{\rho _*(x)}} \int (|\nabla |\rho _*)(x)\, dx \, .
\end{equation}
If we equate $0$, otherwise arbitrary  integration constant,    we have  associated with $\rho _*(x)= 2/\pi (1+x^2)^2$ an
 admissible drift function:
\begin{equation}
b(x)= -{\frac{\gamma }8}\,  (x^3 +  3 x)  \, .
\end{equation}
Thus, if we wish to deal with the Langevin process  associated with the quadratic Cauchy  density (40),
the proper drift form is  given above.

\subsection{"Topological" Cauchy family}

We may  consider  various  probability densities  as  trial ones. Let us  pay attention to a broader class of densities
 that bear  close  affinity  with the   Cauchy noise.
With a given continuous probability distribution $\rho $  we associate  its Shannon entropy
 $S(\rho ) = -\int \rho \, \ln \rho \, dx$, \cite{kapur}.
  If an expectation value $E[\ln (1+x^2)]$ is prescribed (e.g. fixed),
  the maximum entropy probability function belongs to a one-parameter family
 \begin{equation}
\rho _* (x)= {\frac{\Gamma (\alpha )}{\sqrt{\pi } \Gamma (\alpha -1/2))}}\, {\frac{1}{(1+x^2)^{\alpha }}}
 \end{equation}
where $\alpha  >1/2$, \cite{kapur}.

 The gamma function $\Gamma (\alpha ) = \int_0^{\infty } \exp(-t)\, t^{\alpha -1}\, dt $ we specialize to integer
$\alpha = n+1$-values, with $n\geq 0$. Then $\Gamma (n+1) = n!$  and $\Gamma (\alpha -1/2) \rightarrow
\Gamma (n+ 1/2)= [(2n)!\sqrt{\pi }]/ n!2^{2n}$.

The Cauchy distribution is a special case of the above $\rho _*$ that corresponds to $\alpha =1$.
The density (40) is the second, $\alpha =2$,
 member of the $\alpha $-integer hierarchy (we presume  $\sigma =1$).

To elucidate an intimate connection with the underlying  Cauchy noise, let us invoke the Fourier transform ${\cal{F}}[\rho _*]$,
 e.g. a characteristic function of $\rho _*$.
Namely, we have, \cite{stephanovich}:
\begin{equation}
{\cal{F}}[1/(x^2  + \gamma ^2)^n] = {\frac{(-1)^{n-1}\, \pi }{(n-1)!}}\,
  {\frac{\partial ^{n-1}}{\partial z^{n-1} }}  \left[ {\frac{\exp(-\sqrt{z} |p|)}{\sqrt{z}}} \right]_{z=\gamma ^2} \, .
\end{equation}
For the (unnormalized, $\alpha =1$)   Cauchy  density  $1/(x^2 + \gamma ^2)$,    we infer $(\pi /\gamma )\exp(-\gamma |p|)$.
For the quadratic ($\alpha =2 $) case, we obtain $(\pi / 2\gamma ^2)[ |p| + (1/\gamma )] \exp(-\gamma |p|)$.

 Given an invariant probability density of the form  (47). Assume that   ${\cal{V}}(x)$ can be inferred, e.g. exists,   and  additionally
   fits  the restrictions of Corollary 2 in  Ref. \cite{olk}.

   Then, we  can  be sure that the corresponding
   "topological" Cauchy-type process can  be consistently  defined.
For each concrete   ${\cal{V}}(x)$,  the resulting Markov  stochastic process  of the jump-type,
 that  is determined by the Cauchy generator  plus a suitable potential function, appears to be  unique.
Specifically, let us consider
\begin{equation}
\rho _* (x) = {\frac{16}{5\pi }} \, {\frac{1}{(1+x^2)^4}} \, .
\end{equation}
By evaluating   principal value integrals,
  we end up with the following expression for the  Feynman-Kac (semigroup) potential:
\begin{equation}
{\cal{V}}(x)= {\frac{\gamma }{2(1+x^2)}}\,  (x^4 +6x^2 -3)\, .
\end{equation}

The potential is bounded from below, its minimum at $x=0$ equals $-3\gamma /2$.  For large values of $|x|$, the potential behaves as
$\sim (\gamma /2) x^2$ i.e. shows up a harmonic behavior.

Apart from  the unbounded-ness  of ${\cal{V}}(x)$ from above, this potential
obeys the minimal requirements of Corollary 2 in Ref. \cite{olk}: can be made positive (add a suitable constant),
 is  locally bounded (e.g. is bounded  on  each compact set) and is measurable (e.g.   can be arbitrarily well
  approximated by means of sequences of step functions).

We can readily address the problem (vi) o Section 3.2.3. Be employing  the density (49), we  get,
\cite{stephanovich}:
\begin{equation}
b(x) = - {\frac{\gamma x}{16}}\, (5x^6 + 21x^4 + 35 x^2 +35)\, .
\end{equation}
This a bit discouraging  expression shows  a linear friction  $b\sim - x$  for small $x$ and a strong taming behavior
 $b\sim -x^7$ for large $x$.

 \section{Conclusions and prospects}

  Insightful, explicitly  solvable  models  are scarce in   theoretical studies of L\'{e}vy flights, in
 the presence of external potentials  and/or  external conservative  forces. Therefore, our major task was
 to find novel   examples,  that would shed some  light on apparent discrepancies between dynamical patterns of
  behavior   associated   with   two  different   fractional  transport   equations (29) and (31) that  are met in the literature.

  We have demonstrated that  so-called topological L\'{e}vy processes form a subclass  of solutions to the  Schr\"{o}dinger  boundary
data problem. The pertinent dynamical behavior stems form a suitable L\'{e}vy-Schr\"{o}dinger semigroup. The crucial role of the involved
Feynman-Kac potential has been identified. We have deduced  these potential functions explicitly in a number of cases.

The major gain of those observations is that a mathematical theory of Ref. \cite{olk} tells  one  what are the necessary functional
   properties of admissible Feynman-Kac potentials. Their proper  choice makes a topological L\'{e}vy process a well behaved mathematical
   construct, with a well defined Markovian dynamics and stationary density.
That gives  clear  indications towards any explicit computer-assisted modeling.

Our focus was  upon confinement mechanisms that tame L\'{e}vy flights to the extent  that second moments of their  probability
densities are admissible. We have proved that the  pertinent  dynamical patterns of behavior   stay in close  affinity  in the
 near-equilibrium regime and, in each considered case,    admit common  for both stationary  density.
  In turn, this density determines a functional form of the above mentioned, dynamical  semigroup-defining,  potential function.

Our  statement of  problems involved can be found in Section 3.2.3. We have analyzed  problems (i)-(iii) and  (vi).
A number of  important  and interesting  topics  has been left aside in the present paper.   Namely,  one may   choose  a priori a
simple,  physics motivated, functional form of the  potential ${\cal{V}}$ (like e.g. ${\cal{V}}\sim  x^2$ or $\sim x^4$)
  and thence  the related   L\'{e}vy  and  specifically Cauchy  semigroup.
 The  major technical  problem (to be circumvented by a computer-assisted exploration of the issue) is
 to deduce an explicit functional form of  the corresponding asymptotic invariant density and to  visualize dynamical patterns of
   behavior in the vicinity  a stationary state. This topic has not received attention in papers on topological  super-diffusions
    \cite{brockmann}-\cite{geisel}.

Since, the Schr\"{o}dinger boundary data problem allows for a construction of an  interpolating Markovian processes between any two
  a priori prescribed probability densities, it is  of interest to fix an initial density and choose an invariant density as  an asymptotic (terminal)
  datum. A comparison of a detailed temporal behavior  of the Langevin-based and affine topological process (both sharing the chosen invariant density)
  is of interest. This issue is postponed to the future publication.

 {\bf Acknowledgement:} Partial support from the Laboratory for
Physical Foundations of  Information Processing is gratefully
acknowledged.  I would like to express my gratitude to  Professor Vladimir Stephanovich who detected an error
 in the previous version of Section 4.2 and  cross-checked  a number of  formulas in the paper.

\end{document}